\documentclass[aps,prl,reprint,preprintnumbers,showpacs,floatfix,nofootinbib,superscript address,longbibliography]{revtex4-2}
\usepackage[utf8]{inputenc}
\usepackage{amssymb}
\usepackage{hhline}
\usepackage{amsmath}
\usepackage{mathtools}
\usepackage[dvipsnames]{xcolor}
\usepackage{multirow,tabularx}
\usepackage{graphicx}
\usepackage{natbib}

\usepackage{xspace}
\usepackage{xstring}
\usepackage{titlesec}
\usepackage{parskip}
\allowdisplaybreaks
\usepackage{braket}

\newcommand*\diff{\mathrm{d}}
\newrobustcmd{\pea}[1]{%
	\emph{#1}\textbf{\ \ \ ---}
}
\titleformat{\paragraph}[runin]{\normalfont\normalsize\bfseries}{\emph\theparagraph}{1em}{\pea}

\newcommand*{\ie}{i.e.\@\xspace}
\newcommand*{\eg}{e.g.\@\xspace}
\newcommand*{\cf}{c.f.\@\xspace}

\newcommand*{\eq}{eq.\@\xspace}
\newcommand*{\eqs}{eqs.\@\xspace}

\newcommand*{\rhs}{r.h.s.\@\xspace}

\usepackage{hyperref}
\hypersetup{%
     colorlinks = true,%
     linkcolor = Blue,%
     citecolor = Blue,%
     filecolor = Blue,%
     urlcolor = Blue%
     }%

\newcommand{\gs}{\Upsilon}

\newcommand{\cs}{\tilde{c}_{\text{sph}}}

\begin{document}

\title{No Warm Inflation From a Vanilla Axion}

\author{Sebastian Zell}
\email{sebastian.zell@lmu.de}
\affiliation{Arnold Sommerfeld Center, Ludwig-Maximilians-Universit\"at, Theresienstraße 37, 80333 M\"unchen, Germany}
\affiliation{Max-Planck-Institut für Physik, Boltzmannstr. 8, 85748 Garching b.\ M\"unchen, Germany}
\affiliation{Centre for Cosmology, Particle Physics and Phenomenology -- CP3, Universit\'e catholique de Louvain, B-1348 Louvain-la-Neuve, Belgium}


\begin{abstract}
	At finite temperature, the coupling of an axion to non-Abelian gauge fields causes dissipation due to sphaleron heating. This mechanism has been considered as ideal for realizing warm inflation since it can lead to large thermal friction while preserving the flatness of the potential. We show, however, that requiring standard properties of an axion -- in particular a discrete shift symmetry -- excludes the strong regime of warm slow-roll inflation, in which thermal friction dominates. The present argument, which does not rely on any phenomenological input, leaves room for the weak regime of warm axion inflation, but in this case a super-Planckian decay constant represents a well-known issue. Finally, we discuss non-minimal and axion-like models as way out.
\end{abstract}

\maketitle

\paragraph*{Cold vs.\@\xspace warm inflation} There is overwhelming evidence that a hot Big Bang (BB) took place about 14 billion years ago. At this point, our Universe was small, dense, and filled with a nearly homogeneous plasma. Subsequently, it expanded and cooled, and the initial inhomogeneities developed into the structures that we see today, such as galaxies, stars and planets. This leads to a fundamental question about the cause of the BB.
Arguably, the leading candidate to address it is inflation \cite{Starobinsky:1980te, Guth:1980zm, Linde:1981mu, Mukhanov:1981xt} -- a phase of accelerated expansion preceding the BB. If it lasts long enough, corresponding to an amplification of space by a factor of at least $\sim \text{e}^{50}$, it can set the initial conditions for the BB while accounting for the observed homogeneity on macroscopic scales. 
The paradigm of inflation leads to predictions that are in excellent agreement with cosmological observations, in particular of the cosmic microwave background (CMB) \cite{Planck:2018jri, BICEP:2021xfz}. 

The simplest way to realize inflation consists in endowing a scalar field $\varphi$ -- the so-called inflaton -- with a large potential energy $V$ (see \cite{Riotto:2002yw,Mukhanov:2005sc,Gorbunov:2011zzc,Baumann:2022mni} for reviews). Key for maintaining accelerated expansion over an extended period of time and for achieving at least $50$ e-folds of spatial amplification is a source of friction that slows down the movement of the inflaton along its potential. The original proposals \cite{Starobinsky:1980te, Guth:1980zm, Linde:1981mu, Mukhanov:1981xt} relied on Hubble friction, \ie the fact that $\varphi$ is decelerated in the background of an expanding universe. This effect is only efficient if the potential is sufficiently flat (as measured in units of the Planck mass $M_P$). However, such potentials are regarded as problematic since it has proven difficult to maintain flatness in the presence of quantum corrections (see \cite{Copeland:1994vg,Binetruy:1996xj,Dvali:1998pa,Arkani-Hamed:2003wrq,Baumann:2014nda}).

Addressing this criticism against inflation, a second possibility for achieving slow-roll was developed: In \emph{warm inflation}, the main source of friction is the interaction with a thermal bath \cite{Berera:1995wh,Berera:1995ie}, which allows for inflationary expansion on much steeper potentials. However, warm inflation comes with its own challenges. The same effect that generates a sufficiently large thermal friction generically also causes strong corrections to the potential, again spoiling inflation \cite{Yokoyama:1998ju}.

The leading solution to this issue is provided by using the axion of a non-Abelian gauge group as inflaton.\footnote
{As originally proposed \cite{Freese:1990rb}, this idea of natural inflation can also be realized at zero temperature.}
In this case, sphalerons \cite{Kuzmin:1985mm,McLerran:1990de} lead to an effect dubbed \emph{sphaleron heating} that is able to induce a large thermal friction without causing relevant changes in the potential \cite{Berghaus:2019whh,Laine:2021ego,Mirbabayi:2022cbt,Klose:2022rxh,Kolesova:2023yfp}. Therefore, sphaleron heating has come to be considered as ideal mechanism for driving warm inflation \cite{Berghaus:2019whh,Laine:2021ego,Klose:2022knn,Mirbabayi:2022cbt,Klose:2022rxh,Kolesova:2023yfp,Drewes:2023khq,Berghaus:2024zfg}.
However, difficulties have already been observed in reconciling warm axion inflation with standard properties of an axion, especially its discrete shift symmetry \cite{Klose:2022rxh}.
 \emph{This motivates us to show that demanding friction to be dominated by sphaleron heating together with the requirement of preserving the usual properties of a ``vanilla'' axion in fact excludes slow-roll inflation.} 

First, we shall review warm axion inflation driven by sphaleron heating, before the class of theories under consideration and our main statement about the incompatibility with slow-roll inflation will be made precise starting from  \eq \eqref{generalTheory}. The expert reader may choose to skip forward to this point. After the subsequent simple proof of our claim culminating in \eq \eqref{result}, we will show that the present argument does not exclude the so-called weak regime of warm inflation, in which thermal friction does not dominate. In this case, however, a super-Planckian decay constant represents a well-known issue. Finally, we will comment on ``non-vanilla'' (axion-like) models and directions forward.

\paragraph*{Vanilla scenario of warm axion inflation}
Once effects of a finite temperature $T$ are included, the equation of motion of the inflaton is (with dot and prime denoting derivatives with respect to time and $\varphi$, respectively, and recent reviews in \cite{Kamali:2023lzq,Berera:2023liv})
\begin{equation} \label{EOM}
	  V' + (3 H + \gs(T)) \dot{\varphi} = -\ddot{\varphi} \;,
\end{equation}
where $H$ is the Hubble scale while $\gs(T)$ corresponds to the coefficient of thermal dissipation. Defining
\begin{equation} \label{Q}
	Q \equiv \frac{\gs(T)}{3 H} \;,
\end{equation}
we see that the main source of friction is thermal if $Q>1$ -- this is the so-called \emph{strong regime} of warm inflation \cite{Laine:2021ego} on which we shall focus. It is important to remark that a finite temperature can be relevant even if $Q<1$: As long as $T\gtrsim H$, perturbations are strongly influenced by thermal effects, which corresponds to the \emph{weak regime} of warm inflation \cite{Laine:2021ego}.
For all temperatures, friction-dominated motion, leading to $\ddot{\varphi} \approx 0$ and accelerated cosmological expansion, is achieved as long as the first slow-roll parameter,
\begin{equation} \label{epsilon}
	\epsilon \equiv \frac{1}{1+Q} \frac{M_P^2 V'^2}{2 V^2} \;,
\end{equation}
is smaller than $1$. Clearly, thermal effects with $Q\gg 1$ allow for inflation on much steeper potentials as opposed to the cold regime.

The axion, which was originally introduced for a dynamical mechanism to address the strong CP-problem of QCD \cite{Peccei:1977ur,Peccei:1977hh,Wilczek:1977pj,Weinberg:1977ma} (see \cite{DiLuzio:2020wdo} for a review), is defined by its coupling to the topological density of a non-Abelian gauge group $SU(N_c)$:
\begin{equation}\label{axionCoupling}
	\frac{\varphi}{f}
	\frac{\alpha\, \text{Tr}\, G_{\mu\nu} \tilde{G}^{\mu\nu}}{4 \pi} \equiv
	\frac{\varphi}{f} q \;.
\end{equation}
Here $f$ is the decay constant, $\alpha$ represents the gauge coupling, and $G_{\mu\nu}$ and  $\tilde{G}^{\mu\nu}$ are the field strength of the gauge group and its dual, respectively.
 The vacuum structure of the $SU(N_c)$ gauge group is non-trivial \cite{Callan:1976je,Jackiw:1976pf}, and for sufficiently large $T$, sphalerons mediate unsuppressed transitions between different vacua \cite{Kuzmin:1985mm}. In this background, the thermal average in the low-frequency regime yields \cite{McLerran:1990de,Arnold:1996dy}
\begin{equation} \label{sphaleronRate}
	-  q  =   \frac{\Gamma_{\text{sph}}}{T}\frac{\dot{\varphi}}{f}  \;, \qquad \Gamma_{\text{sph}} =  \cs T^4 \;.
\end{equation}
Here $\Gamma_{\text{sph}}$ is the sphaleron rate and its prefactor is suppressed as $\cs \sim (\alpha N_c)^5$, where we did not display a factor of order $1$ that depends weakly on other model parameters and dynamical quantities (see \cite{Moore:2010jd,Laine:2021ego,Laine:2022ytc,Klose:2022rxh}).
Plugging \eq \eqref{sphaleronRate} into the equation of motion following from \eq \eqref{axionCoupling}, we see that the resulting friction rate is \cite{McLerran:1990de,Arnold:1996dy}
\begin{equation} \label{sphaleronHeating}
	\gs(T) = \frac{\Gamma_{\text{sph}}}{f^2 T} \;,
\end{equation}
corresponding to a process dubbed \emph{sphaleron heating} \cite{Mirbabayi:2022cbt}. Crucially, the fact that $ \text{Tr}\, G_{\mu\nu} \tilde{G}^{\mu\nu}$ is a total derivative implies that the coupling \eqref{axionCoupling} is shift-symmetric and so it cannot cause perturbative corrections to the flatness of the potential. Therefore, warm axion inflation is protected from dangerous quantum effects \cite{Berghaus:2019whh,Laine:2021ego,Mirbabayi:2022cbt,Klose:2022rxh,Kolesova:2023yfp}, which resolves the problem raised in \cite{Yokoyama:1998ju}.\footnote
{In the absence of a shift-symmetry, a large dissipation coefficient $\gs(T)$ is almost inevitably accompanied by 
sizable finite-temperature contributions to the effective potential since both effects come from the imaginary and real part of the same diagram \cite{Buldgen:2019dus, Kainulainen:2021eki}.}

At the non-perturbative level (and independently of thermal effects), a coupling of the form \eqref{axionCoupling} can induce a potential for the axion \cite{Peccei:1977ur,Peccei:1977hh,Wilczek:1977pj,Weinberg:1977ma}, thereby breaking the continuous shift invariance. However, the topological character of $G_{\mu\nu} \tilde{G}^{\mu\nu}$ ensures that a discrete shift symmetry is still preserved,
\begin{equation} \label{shiftSymmetryAxion}
	\varphi \rightarrow \varphi + 2 \pi f \;.
\end{equation}
This invariance represents a key property of the axion. First, it strongly constrains the form of the axion potential, \eg by excluding simple polynomial contributions, thereby selecting the operator \eqref{axionCoupling} as leading interaction. This is crucial for the predictivity of axion models and for distinguishing them from generic inflationary scenarios with a priori arbitrary potentials.
Second, conventional axion models such as \cite{Kim:1979if,Shifman:1979if,Zhitnitsky:1980tq,Dine:1981rt} identify $\varphi/f$ as the phase of a complex PQ-field, and so the shift symmetry \eqref{shiftSymmetryAxion} must be obeyed by construction.

Below the confinement scale $\Lambda \equiv \sqrt{m f}$, the non-perturbatively generated periodic potential can be approximated as \cite{GrillidiCortona:2015jxo}\footnote
{An alternative, less accurate  approximation is (see \cite{Freese:1990rb,Marsh:2015xka})
\begin{equation}  \label{VAxionAlternative}
		V(\varphi) = m^2 f^2 \left(1 - \cos\left(\frac{\varphi}{f}\right)\right)	 \;,
\end{equation}
which agrees with \eq \eqref{VAxion} up to order $\varphi^2$. The difference between \eqs \eqref{VAxionAlternative} and \eqref{VAxion} is inessential for the present discussion.}
\begin{equation} \label{VAxion}
	V(\varphi) =  4 m^2 f^2 \left(1 - \left|\cos\left(\frac{\varphi}{2 f}\right)\right|\right)
 \;,
\end{equation}
where $m$ is the mass of the axion. Evidently, inflation is impossible with simple QCD axion models since $\Lambda_{\text{QCD}} \sim 100\, \text{MeV}$ certainly lies below inflationary energies and moreover the presence of light quarks is known to suppress sphaleron heating \cite{McLerran:1990de,Berghaus:2019whh,Berghaus:2020ekh}, although it is straightforward to construct models that feature sphaleron heating in the presence of light fermions \cite{Drewes:2023khq}. Instead, \eq \eqref{VAxion} with parameters suitable for inflation can \eg arise from confinement above inflationary scales with an unbroken subgroup $SU(N_c)$ \cite{Klose:2022rxh} or from heavy axion models coupled to an additional dark gauge group (see \cite{Tye:1981zy,Rubakov:1997vp,Berezhiani:2000gh,Hook:2014cda,Fukuda:2015ana,Dimopoulos:2016lvn,Hook:2019qoh,Valenti:2022tsc,Co:2022bqq,Dunsky:2023ucb} and overview in \cite{Berghaus:2024zfg}). 
Realizations of warm axion inflation involving dark gauge groups have been discussed in \cite{Mohanty:2008ab,Mishra:2011vh,Visinelli:2011jy,Ferreira:2017lnd,Ferreira:2017wlx,Kamali:2019ppi,Berghaus:2019whh}, with a concrete implementation of sphaleron heating in \cite{Berghaus:2019whh}, while a model of a heavy QCD axion coupled to additional particles was proposed recently \cite{Berghaus:2024zfg}. 

So far, we have not considered phenomenology. In the strong regime $Q>1$, it is well-known that periodic potentials of the form \eqref{VAxion} tend to produce a blue scalar spectral index \cite{Berghaus:2019whh,Montefalcone:2022jfw}, which does not match the experimentally observed red tilt. Both in \cite{Berghaus:2019whh} and \cite{Berghaus:2024zfg}, consistency with observations has been achieved in a hybrid scenario \cite{Linde:1991km,Linde:1993cn} of inflation. 
 
 \paragraph*{Main statement} Leaving aside phenomenology, the above discussions imply that warm inflation driven by sphaleron heating with a vanilla axion is described the equation of motion
 \begin{align} \label{generalTheory}
   &V' + \gs(T) \dot{\varphi} \approx 0 \;,
   \intertext{where the friction coefficient and potential obey the properties}
    \ &\gs(T) = \frac{\cs T^3}{f^2} \;, \label{Y} \\
\ & V(\varphi) \equiv m^2 f^2 \mathcal{V}(\varphi/f) \;,\label{V} \\
         \ &\mathcal{V}(\varphi/f + 2 \pi) = \mathcal{V}(\varphi/f) \;, \label{shift}
 \end{align}
with $\mathcal{V}(\varphi/f)\lesssim 1$. Evidently, the above form of $\gs(T)$ can be derived from sphaleron heating (\cf \eq \eqref{sphaleronRate}) and \eq \eqref{shift} implements the crucial discrete shift symmetry \eqref{shiftSymmetryAxion}.
 Matching the non-perturbative axion potential \eqref{VAxion}, we additionally assume $\mathcal{V}=0$ and $\mathcal{V''}\sim 1$ at the minimum of the potential so that the vacuum mass of $\varphi$ is on the order of $m$, but this property is not essential for our argument. 
 Although the requirements just described do not appear to be very constraining, we shall show that in fact they do not allow for slow-roll inflation, \ie \emph{sphaleron heating with a vanilla axion is unable to drive the strong regime of warm inflation.}
 
 It is important to point out that the same decay constant $f$ appears both in the friction rate \eqref{Y} and the potential \eqref{V} (see also \cite{Klose:2022knn,Klose:2022rxh}). If we momentarily considered a different $f_V$ in the potential, then $f_V>f$ is excluded since a shift symmetry $\mathcal{V}(\varphi/f_V + 2 \pi) = \mathcal{V}(\varphi/f_V)$ would not suffice for preserving invariance \eqref{shiftSymmetryAxion} under $\varphi \rightarrow \varphi + 2 \pi f$. The opposite case $f_V<f$ may be admissible (provided $f/f_V$ is integer), but this scenario is equivalent to a single decay constant $f_V$ together with a smaller value of $\cs$, and so our argument still applies.

We emphasize that our assumptions, as detailed in \eqs \eqref{generalTheory}--\eqref{shift}, cover a wide range of scenarios. First, we do not assume that the gauge group is vanilla, thereby including among others scenarios of partial confinement \cite{Klose:2022rxh}, modified axion potentials  \cite{Klose:2022rxh} and heavy axion-models \cite{Tye:1981zy,Rubakov:1997vp,Berezhiani:2000gh,Hook:2014cda,Fukuda:2015ana,Dimopoulos:2016lvn,Hook:2019qoh,Valenti:2022tsc,Co:2022bqq,Dunsky:2023ucb}. Moreover, if the axion arises as phase of a complex PQ-field, our conclusion still applies even if PQ-symmetry is broken explicitly.\footnote
{For example, gravity is expected to violate PQ-symmetry by operators of the form $\left(\exp(i\varphi/f)\right)^n + \text{h.c.}$, which are suppressed by powers of $M_P$ \cite{Kamionkowski:1992mf} (see \eg also \cite{Alvey:2020nyh,Catinari:2024zon}). These generate additional contributions to the axion potential \cite{Kamionkowski:1992mf,Alvey:2020nyh,Catinari:2024zon}:
	\begin{equation}
		\Delta V \sim \tilde{\Lambda}^4 \left(1 - \cos\left(n \frac{\varphi}{f} + \delta\right)\right) \;,
	\end{equation}
	where $\tilde{\Lambda}$ is an energy scale and $\delta$ represents a phase. Evidently, the modified potential $V+\Delta V$ still obeys the discrete shift symmetry \eqref{shiftSymmetryAxion}. The same is true if the explicit breaking of PQ-symmetry happens through a non-minimal coupling to gravity \cite{Takahashi:2015waa,Berbig:2024ufe}.}

For the following straightforward argument, we shall consider \eqs \eqref{generalTheory}--\eqref{shift} in the strong warm regime of slow-roll inflation, \ie for $Q > 1$ and $\epsilon < 1$. We use the corresponding standard approximations, namely slow-roll in \eq \eqref{generalTheory}, a Hubble scale dominated by the potential, $H \approx \sqrt{V/3}/M_P$, and moreover neglect temperature-dependent contributions to the energy \cite{Berghaus:2019whh,Mirbabayi:2022cbt,Kamali:2023lzq,Drewes:2023khq,Berghaus:2024zfg}.
 Our goal is to show this cannot lead to a sufficiently long period of inflation. As is well-known, stationary sphaleron heating determines the energy density of the radiation bath to be $g_\star T^4 \pi^2/30 = \gs(T) \dot{\varphi}^2/(4H)$, where $g_\star$ is the effective number of degrees of freedom, so that we can express temperature as (see \cite{Berghaus:2019whh,Drewes:2023khq})
\begin{equation} \label{T}
T=	\left(\frac{15 \sqrt{3} f^2 M_p  V^{'2}}{2 \pi ^2 g_\star \cs\sqrt{V}} \right)^{1/7} \;.
\end{equation}
Plugging this alongside $\gs(T)$ into \eq \eqref{Q}, we get
\begin{equation} \label{QEv}
Q = \left(\frac{375\, \cs^4 M_p^{10} V^{'6} }{8 \pi^6 g_\star^3 f^8 V^5}\right)^{1/7} \;.
\end{equation}
Now demanding $Q>1$ leads to a lower bound on the derivative of the potential:
\begin{equation} \label{VpLower}
V'>	\left(\frac{8 \pi^6 g_\star^3 f^8 V^5}{375\, \cs^4 M_p^{10}}\right)^{1/6} \;.
\end{equation}
Only if the potential is sufficiently steep, it can generate enough thermal friction to maintain the strong regime of warm inflation.
On the other hand, inserting \eq \eqref{QEv} into \eq \eqref{epsilon} shows that the slow-roll parameter scales as $\epsilon \sim V^{'8/7}$, and so requiring $\epsilon<1$ yields an upper bound on $V'$. The potential must not be too steep since otherwise slow roll is violated. It is the combination of these lower and upper bounds on $V'$ that cannot be fulfilled for a sufficiently long period of time.

In order to see this, we compute the number of e-folds as usual (see \eg \cite{Kamali:2023lzq,Berghaus:2024zfg}):
\begin{align}
	N &= \frac{1}{M_P^2} \int_{\varphi_{\text{end}}}^{\varphi_{\text{in}}} \diff \varphi\, \frac{V}{V'} (1 + Q) \nonumber\\
	& < \frac{2}{M_P^2} \int_{\varphi_{\text{end}}}^{\varphi_{\text{in}}}  \diff \varphi\,  \left(\frac{375\, \cs^4 M_p^{10} V^2 }{8 \pi^6 g_\star^3 f^8 V^{'}}\right)^{1/7} \;,
\intertext{where $\varphi_{\text{in}}$ and $\varphi_{\text{end}}$ are initial and final field values, respectively, and we used \eq \eqref{QEv} in the second step (along with $Q+1< 2Q$). Now we come to the key assumptions about our model. First, the lower bound \eqref{VpLower} on $V'$ resulting from maintaining the strong regime of warm inflation together with the monotonicity of $V$ imply}
	N & < \left(\varphi_{\text{in}} - \varphi_{\text{end}}\right) \left(\frac{3000\, \cs^4 V(\varphi_{\text{in}})}{\pi^6 g_\star^3 f^8 M_p^2}\right)^{1/6} \;.
\end{align}
Second, it follows from periodicity \eqref{shift} that the field range is compact, $\varphi_{\text{in}} - \varphi_{\text{end}} < 2 \pi f$, and so we arrive at
\begin{equation}
	N  < 2\left(\frac{3000\, \cs^4 V(\varphi_{\text{in}})}{g_\star^3 f^2 M_p^2}\right)^{1/6} 
	 < 7.6\frac{\cs^{2/3}}{g_\star^{1/2}} \left(\frac{m}{M_p}\right)^{1/3}    , \label{result}
\end{equation}
 where we additionally plugged in the form \eqref{V} of the potential in the last step. The ratio of dimensionless quantities in the first fraction works to suppress $N$ since $\cs\lesssim 1$ (see \cite{Laine:2021ego,Laine:2022ytc}) and $g_\star\geq 1$ by definition.  Clearly, the mass of any particle must obey $m<M_P$ and 
so \eq \eqref{result} shows that it is impossible to achieve a sufficient number of e-foldings. This proves the main statement of the paper. We emphasize that the bound \eqref{result} is very conservative -- \eg the first maximum of the axion potential generically occurs for $\varphi < \pi f$ (see \eq \eqref{VAxion}), which introduces another factor of $1/2$ on the \rhs of \eq \eqref{result} -- and so it appears challenging to construct a viable model of a vanilla axion in which sphaleron heating maintains the strong regime of warm inflation for even a single e-fold.

 \paragraph*{Ways out} Searching for possible ways to still implement inflation driven by thermal friction, one might contemplate a short transient phase of $Q>1$ during the generation of CMB perturbations followed by different inflationary dynamics. However, the scaling $Q\sim V^{'6/7}$ of \eqref{QEv} implies that this generically requires a non-trivial potential where $V'$ is not monotonic (see also \cite{Drewes:2023khq}).
In a similar spirit, the above argument relies on the slow-roll approximation and so \eg a phase of ultra-slow roll could in principle provide a way out. Although this has to be checked on a case-by-case basis, it seems unlikely that such an approach will be successful as ultra-slow roll can at most last for a few e-foldings before invalidating inflationary perturbation theory \cite{Germani:2017bcs} (see also \cite{Poisson:2023tja}) and moreover it seems difficult to maintain a sufficiently high temperature of the thermal plasma \cite{Biswas:2023jcd}.

In \cite{Berghaus:2019whh,Berghaus:2024zfg}, the strong regime of warm inflation is implemented by abandoning the discrete shift symmetry \eqref{shiftSymmetryAxion}.\footnote
{Both \cite{Berghaus:2019whh} and \cite{Berghaus:2024zfg} rely on a hybrid setup, in which inflation can only take place for $\varphi>\varphi_c$, where $\varphi_c$ is a critical field value. In \cite{Berghaus:2019whh}, $\varphi_c > 10^8 f$ and \cite{Berghaus:2024zfg} has $\varphi_c > 10^4 f$ in the strong regime, so in both cases the shift symmetry \eqref{shiftSymmetryAxion} is broken.}
Another option consists in taking $\cs \gg 1$ (see \cite{Reyimuaji:2020bkm,Correa:2022ngq,Montefalcone:2022jfw}), which however leads to a concern about strong coupling.\footnote
{One may try to absorb $\cs \gg 1$ in terms of two different decay constants, defining a small $\tilde{f}=f/\sqrt{\cs}$ in the friction rate \eqref{Y} while still using $f$ in the potential (\eqs \eqref{V} and \eqref{shift}).}
In a PQ-scenario, it is possible to include a non-minimal coupling of the radial field to gravity, which would lead to a decay constant that depends on the inflationary Hubble scale $H$ \cite{Linde:1991km,Higaki:2014ooa,Choi:2014uaa,Chun:2014xva,Fairbairn:2014zta, Ballesteros:2016xej,Rigouzzo:2025hza}. Evidently, one can also replace sphaleron heating by another source of thermal friction with a different $\gs(T)$ (see \cite{Kamali:2023lzq,Ballesteros:2023dno} and references therein), but in this case care must be taken to avoid the problem \cite{Yokoyama:1998ju} of spoiling inflation because of large thermal corrections to the potential. Finally, it is possible to consider an axion-like particle (ALP) that does not necessarily obey a discrete shift symmetry \eqref{shiftSymmetryAxion} or arise as phase of a complex field (see review \cite{Marsh:2015xka} and \eg \cite{Karananas:2024xja} for an independent motivation to consider such an ALP).

 \paragraph*{Weak regime} As already mentioned, a finite temperature can have an important effect even if $Q<1$. In this case, the main source of friction is Hubble dilution, \ie the background evolution of the inflaton field is to leading order unchanged by thermal effects. As long as $T>H$, however, perturbations and therefore inflationary observables are crucially altered by the presence of a thermal plasma \cite{Gupta:2002kn,Hall:2003zp,Gupta:2005nh,Moss:2007cv,Graham:2009,Moss:2011qc,Bastero-Gil:2011rva, LopezNacir:2011kk,Ramos:2013nsa,Bastero-Gil:2014jsa,Bastero-Gil:2014raa,Bastero-Gil:2019rsp,Das:2020xmh,Qiu:2021ytc,Mirbabayi:2022cbt,Klose:2022rxh,Bastero-Gil:2023sub,Kamali:2023lzq,Ballesteros:2023dno,Montefalcone:2023pvh,Drewes:2023oxg,Laine:2024wyv} -- this is the \emph{weak regime} of warm inflation. With an analogous argument as above, requiring $T \gtrsim H$ leads to the condition
 \begin{equation}
 	V' \gtrsim \left(\frac{2 \pi^2 g_\star \cs  V^4}{1215 f^2 M_p^8}\right)^{1/2} \;,
 \end{equation}
 and so the number of e-foldings is bounded as
 \begin{equation}
 	N < \frac{2}{M_P^2} \int_{\varphi_{\text{end}}}^{\varphi_{\text{in}}} \diff \varphi\, \frac{V}{V'} \lesssim \frac{99}{\sqrt{g_\star \cs}} \left(\frac{M_P}{m}\right)^2 \;.
 \end{equation}
Clearly, a restrictive bound on $N$ can now be avoided, especially by choosing the prefactor $\cs$ of thermal friction and the scale of the potential $V$ (equivalently $m$) sufficiently small. While inflation \emph{from} sphaleron heating, corresponding to $Q>1$,  runs into trouble, our argument does not exclude inflation \emph{with} sphaleron heating, \ie $Q<1$. 
	
	When thermal friction is not dominant, slow-roll parameters scale with $(M_P/f)^2$ (\cf \eq \eqref{epsilon}). Therefore, warm axion inflation in the weak regime is only possible with a super-Planckian decay constant $f > M_P$; see \cite{Klose:2022rxh,Kolesova:2023yfp,Laine:2024wyv} for concrete realizations. However, $f > M_P$ puts into question the validity of effective field theory (EFT) \cite{Arkani-Hamed:2003xts,Arkani-Hamed:2003wrq,Banks:2003sx}, and non-vanilla axion models, \eg involving more than one axion field \cite{Kim:2004rp,Dimopoulos:2005ac} or relying on non-trivial string compactifications \cite{Silverstein:2008sg,McAllister:2008hb}, are required in order to evade this concern. As a second independent aspect, the simultaneous presence of sphaleron heating and a non-perturbatively generated potential has so far only been realized with a non-trivial gauge sector (see \cite{Klose:2022rxh,Kolesova:2023yfp,Laine:2024wyv} and discussion above). Therefore, the weak regime of warm axion inflation cannot be viewed as vanilla, either.

 \paragraph*{Summary}
 Among the most compelling virtues of warm inflation is that it permits inflationary expansion on a wider class of potentials, which need not be as flat as in the absence of thermal friction. This addresses one of the central challenges of cold inflation: the difficulty  \cite{Copeland:1994vg,Binetruy:1996xj,Dvali:1998pa,Arkani-Hamed:2003wrq,Baumann:2014nda} of maintaining a flat potential in the presence of quantum corrections. However, when thermal effects dominate, they also tend to introduce large corrections to the inflaton potential, thereby undermining slow-roll conditions \cite{Yokoyama:1998ju}. Sphaleron heating -- arising from coupling an axion to a non-Abelian gauge group at finite temperature -- has been proposed as an elegant solution: the shift symmetry of the axion protects the flatness of the potential even when thermal friction is strong \cite{Berghaus:2019whh,Laine:2021ego,Mirbabayi:2022cbt,Klose:2022rxh,Kolesova:2023yfp}. 
 
 The key insight added by the present work is that sphaleron heating with a vanilla axion, as defined in \eqs \eqref{generalTheory}--\eqref{shift}, cannot support the strong regime of warm inflation. Crucially, this result does not rely on any phenomenological input but follows directly from the fundamental properties of an axion -- most notably, its discrete shift symmetry. Combined with the established fact that axion inflation without strong thermal friction necessitates a super-Planckian decay constant, we conclude that \emph{there is no warm inflation from a vanilla axion}.
  
   \paragraph*{The future of sphaleron heating}
   Sphaleron heating is still relevant in contexts beyond inflation -- \eg for a dynamical explanation of dark energy \cite{Berghaus:2019cls,Berghaus:2020ekh}, where prolonged slow-roll is not essential, or for the generation of dark matter \cite{Papageorgiou:2022prc,Choi:2022nlt,Biondini:2024cpf,McKeen:2024trt} (see also \cite{Freese:2024ogj}). Existing inflationary models employing this mechanism, such as \cite{Berghaus:2019whh,Berghaus:2024zfg}, also remain valid as low-energy EFTs. Crucially, however, their ultraviolet completion must differ from earlier assumptions: the inflaton cannot be an axion, especially not one arising as the phase of a complex scalar field. This has a key implication: in the absence of a discrete shift symmetry,  there is no longer any compelling reason to single out the specific non-renormalizable operator \eqref{axionCoupling}. A much broader range of interactions becomes admissible and a new principle is now required to account for their suppression or absence.
   
  Paradoxically, this apparent limitation opens new avenues for model building. Once freed from the constraints of axion-based EFTs, two important advantages emerge:
   \begin{itemize}
   	\item The EFT of an axion is valid only up to the scale $f$. In contrast, the operator \eqref{axionCoupling} remains under control up to the higher cutoff scale $f/\alpha$.
   	\item The axion potential contains a mass term -- see \eq \eqref{VAxion} -- whereas a generic inflaton may possess a different, potentially more favorable potential.
   \end{itemize}
   In particular, these two features -- the higher effective cutoff and the freedom in potential choice -- enable the construction of new models of warm inflation driven purely by Standard Model interactions, as already demonstrated in \cite{Berghaus:2025dqi}. Such scenarios stand out not only for their minimal field content but also for the possibility to yield observable signals in axion experiments.
   
\begin{acknowledgments}

\paragraph*{Acknowledgments}  This work was started at the Albert Einstein Center (AEC) at the University of Bern and I thank the AEC and especially Mikko Laine for hospitality and support through their visitor program. I am very grateful to Marco Drewes and Mikko Laine for providing initial impulses for this project as well as insightful feedback, and I thank Kim Berghaus and Misha Shaposhnikov for many helpful comments and discussions. Moreover, I am indebted to the participants of the CERN-TH institute ``Particle Production in the Early Universe'' for interesting questions and feedback. This work was supported by the European Research Council Gravites Horizon Grant AO number: 850 173-6 and the \emph{Fonds de la Recherche Scientifique} -- FNRS.

\textbf{Disclaimer:} Funded by the European Union. Views and opinions expressed are however those of
the authors only and do not necessarily reflect those of the European Union or European Research
Council. Neither the European Union nor the granting authority can be held responsible for them.
\end{acknowledgments}

\bibliography{NotINSPIRE,WarmInflation}

\end{document}